\documentclass[aps,prl,twocolumn,floatfix]{revtex4}
\usepackage{amsmath}
\usepackage{subfigure}
\usepackage{graphicx,psfrag}
\usepackage{bm}
\usepackage{multirow}

\begin{document}

\title{Identification of functional information subgraphs in complex networks}

\author{Lu\'is M. A. Bettencourt}
\altaffiliation[Also at the ]{Santa Fe Institute, 1399 Hyde Park Road, Santa Fe NM 87501, USA}
\author{Vadas Gintautas}
\altaffiliation[Also at the ]{Center for Complex Systems Research, Dept. of Physics, University of Illinois at Urbana-Champaign, Urbana IL 61801, USA }
\author{Michael I. Ham}
\altaffiliation[Also at the ]{Center for Network Neuroscience, University of North Texas, Denton TX 76203, USA}
\affiliation{ T-7 and CNLS, Theoretical Division, MS B284 Los Alamos National Laboratory, Los Alamos NM 87545, USA}
 
 \date{\today}

\begin{abstract}
We present a general information theoretic approach for identifying functional subgraphs in complex networks where the dynamics of each node are observable.  We show that the uncertainty in the state of each node can be expressed as a sum of information quantities involving a growing number of correlated variables at other nodes.  We demonstrate that each term in this sum is generated by successively conditioning mutual informations on new measured variables, in a way analogous to a discrete differential calculus. The analogy to a Taylor series suggests efficient search algorithms for determining the state of a target variable in terms of functional groups of other degrees of freedom.  We apply this methodology to electrophysiological recordings of networks of cortical neurons grown {\it in vitro}.  Despite strong stochasticity, we show that each cell's patterns of firing are generally explained by the activity of a small number of other neurons.  We identify these neuronal subgraphs in terms of their mutually redundant or synergetic character and reconstruct neuronal circuits that account for the state of each target cell. 
\end{abstract}

\maketitle
Information plays a central role in conditioning structure and determining collective dynamics in many complex systems.
For example, the ability to process and react to information certainly influences how neurons and synapses, or genes and proteins, interact in large numbers to generate the complexity of cognitive and biological processes. 
Despite their importance, however, systematic methodologies for identifying functional relations between units of successive complexity, involved in information processing and storage, are still largely missing.

Motivated by recent theoretical developments and experimental breakthroughs, new interest has arisen in applications of information theory to dynamical and statistical systems with many degrees of freedom \cite{Borst_1999,Schreiber_2000,Palus_2001}. 
Specifically, it has been shown that information quantities can identify and classify spatial \cite{Bialek_2001} and temporal \cite{Crutchfield_2003} correlations, and reveal if a group of variables may be mutually redundant or synergetic \cite{Schneidman_2003,Bettencourt_2007}. 
In this way an information theoretic treatment of groups of correlated degrees of freedom can reveal their functional roles in terms of arrangements that can serve as memory structures or those capable of processing information.

The application of these insights to identify functional connectivity structure is still just beginning~\cite{Bettencourt_2007} but should provide a useful complement to other established approaches~\cite{Stephan_2000,Milo_2002,Yeger_2004,Ziv_2005} by directly relating observable dynamics or statistics to information structures.
To date, the identification of functional relations between nodes of a complex network has relied on the statistics of {\em motifs}.
These are specific (directed) subgraphs of $k$ nodes that appear more abundantly than expected in randomized networks with the same number of nodes and degree of connectivity  \cite{Stephan_2000,Milo_2002,Yeger_2004}. 
Although powerful for small subgraphs, this approach scales up poorly since the number of different subgraphs explodes combinatorially with increasing number of nodes $k$.
Consequently, the extensive searches that are necessary for measuring motif frequencies become prohibitive beyond about $k\geq 5$.  
A general solution to this curse of dimensionality is to perform targeted searches guided by quantitative expectations for finding the most informative node combinations relative to an external signal or to other parts of the system.

Here we present such an approach, based on the rigorous properties of information theory applied to the correlated statistical state of many variables. 
We show how the uncertainty in the state of any target variable, quantified by its Shannon entropy, can be expressed in terms of a cluster expansion of information quantities involving a successively larger number of variables. 
The sign and magnitude of each term in the expansion determines the functional connectivity among nodes to that order; specifically whether a set of $k$ nodes is functionally independent, redundant, or synergetic.
Because the Shannon entropy is positive definite, this expansion gives a systematic approximation to the state of the target.
As a result the expansion can be truncated at any order and used to construct approximate non-exhaustive search algorithms, analogous to gradient methods in other optimization problems.  
We demonstrate the efficacy of this method through its application to spike time series of cortical neuronal networks grown {\it in vitro}.

Information is a relative quantity, quantifying the increase in predictability (reduction in uncertainty) of a variable's statistical state given knowledge of others with which it is correlated. 
Specifically, the uncertainty in the state of $X$ can be quantified by its Shannon entropy~\cite{Cover_1991} $S(X)= -\sum_{x} p(x) \log_2 p(x)$, where $p(x)$ are the marginals for each state $x$ of $X$.
Note that $S(X)\geq 0$, where $S(X)=0$ corresponds to precise knowledge of $X$ and the probability distribution $p(x)=1$ for some state $x$.  
Measuring correlated variables $Y_i$ to $X$ contributes to knowledge of its state and reduces its uncertainty, thus 
\begin{eqnarray}
    S(X) \geq S(X | \{Y\}_{k-1}) \geq  S(X|\{Y\}_{k}), 
\end{eqnarray}
with $k \leq n$ for $n$ total variables and where $S(X|Y)$ refers to the conditional entropy of $X$ given $Y$~\cite{Cover_1991}.
We use the notation $\{Y\}_{k}$ to refer to the set $Y_1,\ldots,Y_{k}$.
The difference between the entropy of $X$ and its entropy given the joint state of a set $\{Y\}_{k}$ is the information in the set:
\begin{equation}
I(X;\{Y\}_{k})=S(X)-S(X|\{Y\}_{k}) \geq I(X;\{Y\}_{k-1}).
\end{equation}
These relations also specify the optimization problem of minimizing the uncertainty in $X$ given $k$ measurements $\{Y\}_{k}$ within a larger (possibly infinite) set.  
Specifically, if a set exists at some order $k$ so that $S(X|\{Y\}_{k})=0$, and therefore $I(X;\{Y\}_{k})=S(X)$, then it fully determines the state of $X$ and no uncertainty remains.
Each measurement can only reduce or leave unchanged $S(X)$, while information quantities are symmetric under permutation of the $Y_i$, so that the maximal entropy reduction from any given set $\{Y\}_{k}$ is unique. 
The challenge resides in finding the measurement set of size $k$ resulting in the smallest remaining uncertainty. 
The computational complexity of this search grows combinatorially with the number of arrangements of size $k$ within $n$ variables, which quickly becomes prohibitive. 
To evade this problem, we introduce the exact expansion 
\begin{align}
&S(X|\{Y\}_{k}) - S(X) = -I(X;\{Y\}_{k}) \label{eq:generaldeltaS}  \\
& \quad = \sum_{i} \frac{\Delta S(X)}{\Delta Y_i} 
+  \sum_{i>j} \frac{\Delta^2 S(X)}{\Delta Y_i \Delta Y_j} +  \ldots + \frac{\Delta^k S(X)}{\Delta Y_1 \ldots \Delta Y_k}. \nonumber 
\end{align} 
The variational operators in Eq.~\eqref{eq:generaldeltaS} define the change in entropy resulting from a measurement as 
\begin{align}
&\frac{ \Delta S (X)}{\Delta Y_i}\equiv S(X|Y_i)-S(X)= -I(X;Y_i) \\
&\frac{ \Delta^2 S (X)}{\Delta Y_i \Delta Y_j}\equiv\frac{ \Delta}{\Delta Y_i}\biggl[S(X|Y_i)-S(X)\biggr]\\
&\qquad = - \frac{\Delta I (X;Y_i)}{\Delta Y_j}= -I(X;Y_i|Y_j)+I(X;Y_i),
\end{align}
and so on.  
Higher order variations follow automatically from the successive application of the first variation, resulting in a simple chain rule. 
Thus, variations to any order $k$ are symmetrical under permutations of the $Y_i$.
 
\begin{figure}[t]
    \begin{center}
      \includegraphics[width=0.98\columnwidth]{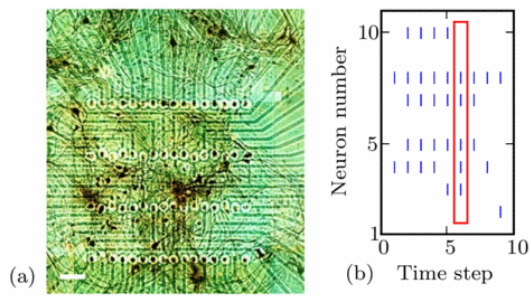}
      \caption{(a) Neuronal culture over a microelectrode array (white circles; bar$=100\mu$m). (b) Detail of a spike timeseries. The box shows network state $011101100$ (bottom to top).\label{fig:f1}}
  \end{center}
\end{figure}
This expansion has two important properties.
First, each term in the expansion at order $k$ accounts for an irreducible set of correlations among a size-$k$ group of $Y_i$ nodes with the target $X$.
Statistical independence among any of the $Y_i$ results in a vanishing contribution to that order and terminates the expansion.
For example, if all $Y_i$ are mutually independent, all variations for $k>1$ vanish identically and the information about $X$ is given by $\sum_i I(X;Y_i)$, that is, the first order terms in Eq.~\eqref{eq:generaldeltaS}.
If the $Y_i$ are correlated in pairs, but not in higher order multiplets, then only terms with $k \leq 2$ will be present, and so on.
Thus, for a system where not all correlations are realized, expression Eq.~\eqref{eq:generaldeltaS} allows the identification of correlated submultiplets, and determines their mutual organization in specifying the state of $X$.

The second important property of this expansion is that the sign of each nonvanishing variation reveals the informational character of the corresponding multiplet.
Specifically, a negative sign indicates that the $k$-multiplet contributes to the state of $X$ with more information than the sum of all its subgroups (synergy), while a positive sign indicates the opposite (redundancy).
We define a synergetic (redundant) core as a set $\{Y\}_{k}$ such that its variation and the variations of all its subgroups of two or more variables are negative (positive).
Explicit examples where the $Y_i$ are inputs of a logical circuit and $X$ is the output (e.g. an AND circuit) confirm that the sign of any variation of the $Y_i$ identifies synergetic arrangements to any order.
Likewise, arrangements where the same information is shared among some of the $Y_i$, as in a Markov chain, result in the sign of the variation indicating redundancy.
Examples of these relations to low orders ($k\leq 3$) have been worked out recently~\cite{Schneidman_2003,Bettencourt_2007}, and their detailed generalization will appear elsewhere \cite{FollowUp}.
We also note that the concept of order-by-order synergy or redundancy captured by each of the terms in Eq.~\eqref{eq:generaldeltaS} generalizes the coefficient of redundancy $R^{S}_{k}(X,\{Y\}_{k})\equiv \sum_{i=1}^k I(X;Y_{i})-I(X;\{Y\}_{k})$ proposed by Schneidman {\it et al.}~\cite{Schneidman_2003}, which refers to the global information deficit (or excess if $R^{S}_{k}<0$) of a multiplet, relative to only the first term in Eq.~\eqref{eq:generaldeltaS}.  

For the remainder of this Letter, we use the expansion in Eq.~\eqref{eq:generaldeltaS} to define the optimization problem of determining the set and decomposition of the $Y_i$ in terms of functional information arrangements that best account for the stochastic behavior of a target $X$.
Because the entropy $S(X|\{Y\}_{k})\geq 0$ for all $k$, this approach defines a well posed optimization problem, with a single global minimum for each set of possible measurements. 

To illustrate this methodology, we apply it to temporal action potential activity from murine frontal cortex neuronal cultures grown {\it in vitro} on non-invasive microelectrode arrays (MEAs)~\cite{Maeda_1995,Keefer_2001,Haldeman_2005}. 
Fig.~\ref{fig:f1}(a) shows an example network growing on an MEA and Fig.~\ref{fig:f1}(b) typical time series data. 
Details of MEA fabrication and culture preparation are described elsewhere~\cite{Gross_1994,Keefer_2001,Bettencourt_2007}. 
These experimental platforms have become model systems for studying living neuronal networks in controlled environments. Recent progress includes studies of dynamical patterns of collective activity \cite{Segev_2002,Beggs_2003,Beggs_2004,Wagenaar_2006,Ham_2007}, connectivity structure \cite{Jia_2004,Bettencourt_2007}, network growth and development \cite{Wagenaar_2006}, and even learning and activity pattern modification \cite{Jimbo_1999,Marom_2002,Demarse_2001} via external stimulation. 
Results presented here refer to $62$ cells of a mature ($42$ days {\it in vitro}; see~\cite{Tateno_2002}) cortical network. 
To analyze patterns of neuronal activity, binary states are constructed [see Fig.~\ref{fig:f1}(b)] for each recorded neuron's time series using temporal bins of $10$ ms; $1$ is recorded if a neuron fires during within a bin and $0$ otherwise.
Probability distributions for states of $k$ neurons are estimated via frequencies and provide the basis for calculating information theoretic quantities.
Probabilities are considered significant if substantially larger than from a null model with randomized spiking at observed rates for each neuron. 
Nearly all of the network activity occurs as global coordinated spiking events, known as network bursts or avalanches \cite{Segev_2001,Beggs_2003,Wagenaar_2006,Ham_2007}.

\begin{figure}[t]
  \begin{center}
      \includegraphics[width=0.98\columnwidth]{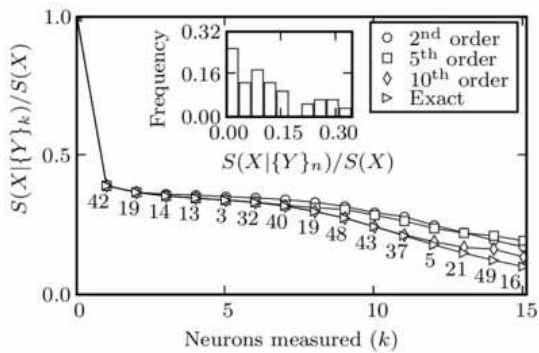}
    \caption{Entropy remaining when other neurons are measured with respect to neuron 46.  Rank is determined by maximizing the variation to various orders.  The neuron numbers appear for the exact curve.  Inset: Histogram of entropy of each neuron remaining after all possible measurements.\label{fig:f2}}
      \end{center}
\end{figure}
Fig.~\ref{fig:f2} shows the relative entropy reduction of a target neuron, due to successive measurements of other neurons.
Different lines correspond to searches for the optimal sequence of measurements at different orders of approximation in the expansion in Eq.~\eqref{eq:generaldeltaS}. 
A search to exact order means that all $I(X;\{Y\}_{k})$ are considered, given the previous $\{Y\}_{k-1}$, and the set $\{Y\}_{k}$ with greatest information gain is chosen.
Most neurons show an initial large drop in entropy due to the measurement of only a few other cells in the network (typically $\leq 5$) and a subsequent slower information gain as more cells are measured. 

\begin{figure}[t]
    \begin{center}
      \includegraphics[width=\columnwidth]{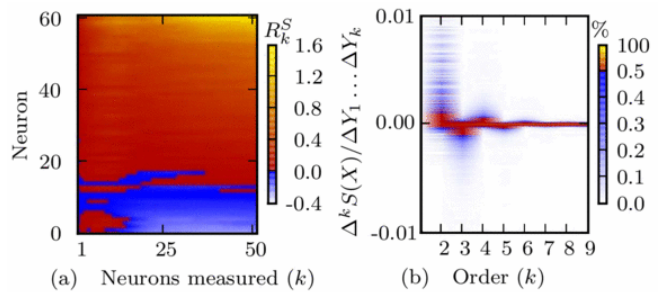} 
      \caption{(a) Sorted global information deficit/excess of a multiplet, relative to the sum of the pairwise mutual informations: $R^S_k$.  (b) Values of each term in the expansion in Eq.~\eqref{eq:generaldeltaS} vs. $k$ for 36,000 randomly sampled variable combinations.  White to blue: $0-0.5\%$; red to yellow: $0.5-100\%$.\label{fig:f3}}
  \end{center}
\end{figure}

Fig.~\ref{fig:f2}(inset) shows the histogram of the ratio of final to initial entropy for all $62$ neurons. 
Final entropy refers to the fraction of a neuron's initial entropy left unaccounted for once the set of all other available neurons is measured. 
Remarkably, the stochastic patterns of most cells can be nearly fully predicted by the activity of others, even if most degrees of freedom in the actual network remain unobserved (we estimate that only about $5-10\%$ of all neurons are measured). 
To better understand the informational nature of arrangements of neurons we show in Fig.~\ref{fig:f3}(a) $R^S_k$ for each of the measured cells in the network. 
By this measure most cell groups are globally redundant (red) relative to their decomposition in terms of purely binary correlations to other cells. 
About a third of the cells, though, show substantial synergy (blue) that persists despite many sequential measurements. 
Fig.~\ref{fig:f3}(b) shows the distribution of each term in the expansion in Eq.~\eqref{eq:generaldeltaS} to order $k$. 
We include all multiplets up to order $k=2$, and thereafter use a random sample of $36,000$ multiplets. 
Recall that the value and sign of each term in the expansion indicates redundancy or synergy relative to the sum of {\it all} submultiplets of lower order.
Globally redundant multiplets often result in terms with alternating signs to lower orders, while a smaller number of multiplets corresponding to synergetic arrangements have negative contributions at every order.  

\begin{figure}[t]
    \begin{center}
      \includegraphics[width=0.98\columnwidth]{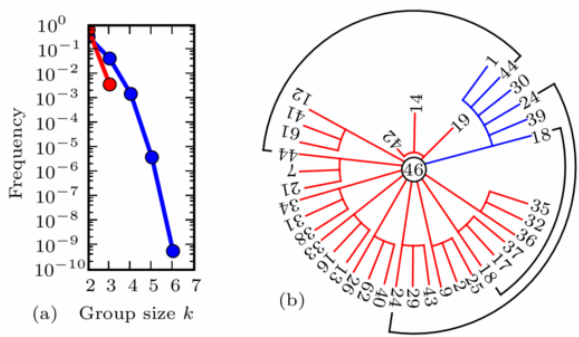}
      \caption{(a) Frequency of redundant (red) and synergetic (blue) cores versus size $k$. (b) Purely redundant (red) and purely synergetic (blue) circuits relative to neuron $46$. Neurons and groups with the most information about $46$ are closest to the center; c.f.~Fig.\ref{fig:f2}. Arcs identify neurons that participate in multiple functional groups.\label{fig:f4}}
  \end{center}
      \end{figure}

Fig.~\ref{fig:f4}(a) shows the frequency of synergetic and redundant cores, while Fig.~\ref{fig:f4}(b) shows the reconstruction of circuits from functional subgraphs which account for the activity of target neuron $46$ of Fig.~\ref{fig:f2}. 
Evidently the target neuron is part of both redundant and synergetic functional multiplets, with the former being substantially more abundant.
The most informative neuron is labeled $42$, but its information about the target is shared to a large extent with neurons $14$ and $19$.
The target neuron is also part of a synergetic circuit with other neurons, several of which are part of smaller mutually redundant subgraphs.
Some of these can, at least partially, be interchanged with other neurons carrying the same information, resulting globally in an interconnected ensemble where specific synergetic functional relationships are embedded on robust redundant cell arrangements.

In summary, we present a new information theoretic approach to constructing functional subgraphs in complex networks where nodes display observable stochastic dynamics.
By performing targeted searches guided by expected information gain from new measurements, we avoid some of the combinatorial issues usually involved in the search for motifs in complex networks.  
We apply this approach to action potential time series from networks of neurons and find that the activity of most neurons is to a large extent determined by the observation of other cells in the network.
This finding is remarkable because only a small portion ($5-10\%$) of cells are accessible to measurement, indicating that large amounts of redundancy characterize neural network dynamics in these cultures.
Although the activity of many neurons can be substantially accounted for by a relatively small number of other cells, an important fraction of a neuron's entropy and detailed firing patterns is contained in multiple cell arrangements of varying size.
These findings agree well with recent neuronal network reconstructions in terms of binary correlations~\cite{Schneidman_2006} and small multiplets~\cite{Bettencourt_2007}, but also provide a new view of the contribution of higher order functional correlations. 
The identification of functional connectivity subgraphs in living neuronal cultures is critical for designing future experiments that promote computational tasks within neural networks, and should find applications more generally in other complex systems. 

We thank G. W. Gross for sharing his extensive expertise with growing and recording the activity of neuronal cultures. We also thank J. Crutchfield, A. Gutfriend, and A. Hagberg for helpful discussions.   This work is supported by LANL's LDRD project 20050411ER.


\begin{thebibliography}{29}
\expandafter\ifx\csname natexlab\endcsname\relax\def\natexlab#1{#1}\fi
\expandafter\ifx\csname bibnamefont\endcsname\relax
  \def\bibnamefont#1{#1}\fi
\expandafter\ifx\csname bibfnamefont\endcsname\relax
  \def\bibfnamefont#1{#1}\fi
\expandafter\ifx\csname citenamefont\endcsname\relax
  \def\citenamefont#1{#1}\fi
\expandafter\ifx\csname url\endcsname\relax
  \def\url#1{\texttt{#1}}\fi
\expandafter\ifx\csname urlprefix\endcsname\relax\def\urlprefix{URL }\fi
\providecommand{\bibinfo}[2]{#2}
\providecommand{\eprint}[2][]{\url{#2}}

\bibitem[{\citenamefont{Borst and Theunissen}(1999)}]{Borst_1999}
\bibinfo{author}{\bibfnamefont{A.}~\bibnamefont{Borst}} \bibnamefont{and}
  \bibinfo{author}{\bibfnamefont{F.~E.} \bibnamefont{Theunissen}},
  \bibinfo{journal}{Nature Neurosci.} \textbf{\bibinfo{volume}{2}},
  \bibinfo{pages}{947} (\bibinfo{year}{1999}).

\bibitem[{\citenamefont{Schreiber}(2000)}]{Schreiber_2000}
\bibinfo{author}{\bibfnamefont{F.}~\bibnamefont{Schreiber}},
  \bibinfo{journal}{Prog. Surf. Sci.} \textbf{\bibinfo{volume}{65}},
  \bibinfo{pages}{151} (\bibinfo{year}{2000}).

\bibitem[{\citenamefont{Palu\ifmmode~\check{s}\else \v{s}\fi{}
  et~al.}(2001)\citenamefont{Palu\ifmmode~\check{s}\else \v{s}\fi{}, Kom\'arek,
  Hrn\ifmmode \check{c}\else \v{c}\fi{}\'i\ifmmode~\check{r}\else \v{r}\fi{},
  and \ifmmode \check{S}\else \v{S}\fi{}t\ifmmode~\check{e}\else
  \v{e}\fi{}rbov\'a}}]{Palus_2001}
\bibinfo{author}{\bibfnamefont{M.}~\bibnamefont{Palu\ifmmode~\check{s}\else
  \v{s}\fi{}}}, \bibinfo{author}{\bibfnamefont{V.}~\bibnamefont{Kom\'arek}},
  \bibinfo{author}{\bibfnamefont{Z.}~\bibnamefont{Hrn\ifmmode \check{c}\else
  \v{c}\fi{}\'i\ifmmode~\check{r}\else \v{r}\fi{}}}, \bibnamefont{and}
  \bibinfo{author}{\bibfnamefont{K.}~\bibnamefont{\ifmmode \check{S}\else
  \v{S}\fi{}t\ifmmode~\check{e}\else \v{e}\fi{}rbov\'a}},
  \bibinfo{journal}{Phys. Rev. E} \textbf{\bibinfo{volume}{63}},
  \bibinfo{pages}{046211} (\bibinfo{year}{2001}).

\bibitem[{\citenamefont{Bialek et~al.}(2001)\citenamefont{Bialek, Nemenman, and
  Tishby}}]{Bialek_2001}
\bibinfo{author}{\bibfnamefont{W.}~\bibnamefont{Bialek}},
  \bibinfo{author}{\bibfnamefont{I.}~\bibnamefont{Nemenman}}, \bibnamefont{and}
  \bibinfo{author}{\bibfnamefont{N.}~\bibnamefont{Tishby}},
  \bibinfo{journal}{Physica A} \textbf{\bibinfo{volume}{302}},
  \bibinfo{pages}{89} (\bibinfo{year}{2001}).

\bibitem[{\citenamefont{Crutchfield and Feldman}(2003)}]{Crutchfield_2003}
\bibinfo{author}{\bibfnamefont{J.~P.} \bibnamefont{Crutchfield}}
  \bibnamefont{and} \bibinfo{author}{\bibfnamefont{D.~P.}
  \bibnamefont{Feldman}}, \bibinfo{journal}{Chaos}
  \textbf{\bibinfo{volume}{15}}, \bibinfo{pages}{25} (\bibinfo{year}{2003}).

\bibitem[{\citenamefont{Schneidman et~al.}(2003)\citenamefont{Schneidman,
  Bialek, and {Berry II}}}]{Schneidman_2003}
\bibinfo{author}{\bibfnamefont{E.}~\bibnamefont{Schneidman}},
  \bibinfo{author}{\bibfnamefont{W.}~\bibnamefont{Bialek}}, \bibnamefont{and}
  \bibinfo{author}{\bibfnamefont{M.~J.} \bibnamefont{{Berry II}}},
  \bibinfo{journal}{J. Neurosci.} \textbf{\bibinfo{volume}{23}},
  \bibinfo{pages}{11539} (\bibinfo{year}{2003}).

\bibitem[{\citenamefont{Bettencourt et~al.}(2007)\citenamefont{Bettencourt,
  Stephens, Ham, and Gross}}]{Bettencourt_2007}
\bibinfo{author}{\bibfnamefont{L.~M.~A.} \bibnamefont{Bettencourt}},
  \bibinfo{author}{\bibfnamefont{G.~J.} \bibnamefont{Stephens}},
  \bibinfo{author}{\bibfnamefont{M.~I.} \bibnamefont{Ham}}, \bibnamefont{and}
  \bibinfo{author}{\bibfnamefont{G.~W.} \bibnamefont{Gross}},
  \bibinfo{journal}{Phys. Rev. E} \textbf{\bibinfo{volume}{75}},
  \bibinfo{pages}{021915} (\bibinfo{year}{2007}).

\bibitem[{\citenamefont{Stephan et~al.}(2000)\citenamefont{Stephan, Hilgetag,
  Burns, O'Neill, Young, and Kotter}}]{Stephan_2000}
\bibinfo{author}{\bibfnamefont{K.~E.} \bibnamefont{Stephan}},
  \bibinfo{author}{\bibfnamefont{C.~C.} \bibnamefont{Hilgetag}},
  \bibinfo{author}{\bibfnamefont{G.~A. P.~C.} \bibnamefont{Burns}},
  \bibinfo{author}{\bibfnamefont{M.~A.} \bibnamefont{O'Neill}},
  \bibinfo{author}{\bibfnamefont{M.~P.} \bibnamefont{Young}}, \bibnamefont{and}
  \bibinfo{author}{\bibfnamefont{R.}~\bibnamefont{Kotter}},
  \bibinfo{journal}{Phil. Trans. R. Soc. Lond. B}
  \textbf{\bibinfo{volume}{355}}, \bibinfo{pages}{111} (\bibinfo{year}{2000}).

\bibitem[{\citenamefont{Milo et~al.}(2002)\citenamefont{Milo, Shen-Orr,
  Itzkovitz, Kashtan, Chklovskii, and Alon}}]{Milo_2002}
\bibinfo{author}{\bibfnamefont{R.}~\bibnamefont{Milo}},
  \bibinfo{author}{\bibfnamefont{S.}~\bibnamefont{Shen-Orr}},
  \bibinfo{author}{\bibfnamefont{S.}~\bibnamefont{Itzkovitz}},
  \bibinfo{author}{\bibfnamefont{N.}~\bibnamefont{Kashtan}},
  \bibinfo{author}{\bibfnamefont{D.}~\bibnamefont{Chklovskii}},
  \bibnamefont{and} \bibinfo{author}{\bibfnamefont{U.}~\bibnamefont{Alon}},
  \bibinfo{journal}{Science} \textbf{\bibinfo{volume}{298}},
  \bibinfo{pages}{824} (\bibinfo{year}{2002}).

\bibitem[{\citenamefont{Yeger-Lotem et~al.}(2004)\citenamefont{Yeger-Lotem,
  Sattath, Kashtan, Itzkovitz, Milo, Pinter, Alon, and Margalit}}]{Yeger_2004}
\bibinfo{author}{\bibfnamefont{E.}~\bibnamefont{Yeger-Lotem}},
  \bibinfo{author}{\bibfnamefont{S.}~\bibnamefont{Sattath}},
  \bibinfo{author}{\bibfnamefont{N.}~\bibnamefont{Kashtan}},
  \bibinfo{author}{\bibfnamefont{S.}~\bibnamefont{Itzkovitz}},
  \bibinfo{author}{\bibfnamefont{R.}~\bibnamefont{Milo}},
  \bibinfo{author}{\bibfnamefont{R.~Y.} \bibnamefont{Pinter}},
  \bibinfo{author}{\bibfnamefont{U.}~\bibnamefont{Alon}}, \bibnamefont{and}
  \bibinfo{author}{\bibfnamefont{H.}~\bibnamefont{Margalit}},
  \bibinfo{journal}{Proc. Natl. Acad. Sci. U.S.A.}
  \textbf{\bibinfo{volume}{101}}, \bibinfo{pages}{5934} (\bibinfo{year}{2004}).

\bibitem[{\citenamefont{Ziv et~al.}(2005)\citenamefont{Ziv, Koytcheff,
  Middendorf, and Wiggins}}]{Ziv_2005}
\bibinfo{author}{\bibfnamefont{E.}~\bibnamefont{Ziv}},
  \bibinfo{author}{\bibfnamefont{R.}~\bibnamefont{Koytcheff}},
  \bibinfo{author}{\bibfnamefont{M.}~\bibnamefont{Middendorf}},
  \bibnamefont{and} \bibinfo{author}{\bibfnamefont{C.}~\bibnamefont{Wiggins}},
  \bibinfo{journal}{Phys. Rev. E} \textbf{\bibinfo{volume}{71}},
  \bibinfo{pages}{016110} (\bibinfo{year}{2005}).

\bibitem[{\citenamefont{Cover and Thomas}(1991)}]{Cover_1991}
\bibinfo{author}{\bibfnamefont{T.~M.} \bibnamefont{Cover}} \bibnamefont{and}
  \bibinfo{author}{\bibfnamefont{J.~A.} \bibnamefont{Thomas}},
  \emph{\bibinfo{title}{Elements of Information Theory}}
  (\bibinfo{publisher}{Wiley}, \bibinfo{address}{New York},
  \bibinfo{year}{1991}).

\bibitem[{\citenamefont{Gintautas et~al.}(2007)\citenamefont{Gintautas,
  Bettencourt, and Ham}}]{FollowUp}
\bibinfo{author}{\bibfnamefont{V.}~\bibnamefont{Gintautas}},
  \bibinfo{author}{\bibfnamefont{L.~M.~A.} \bibnamefont{Bettencourt}},
  \bibnamefont{and} \bibinfo{author}{\bibfnamefont{M.~I.} \bibnamefont{Ham}},
  \bibinfo{journal}{in preparation}  (\bibinfo{year}{2007}).

\bibitem[{\citenamefont{Maeda et~al.}(1995)\citenamefont{Maeda, Robinson, and
  Kawana}}]{Maeda_1995}
\bibinfo{author}{\bibfnamefont{E.}~\bibnamefont{Maeda}},
  \bibinfo{author}{\bibfnamefont{H.~P.} \bibnamefont{Robinson}},
  \bibnamefont{and} \bibinfo{author}{\bibfnamefont{A.}~\bibnamefont{Kawana}},
  \bibinfo{journal}{J. Neurosci.} \textbf{\bibinfo{volume}{15}},
  \bibinfo{pages}{6834} (\bibinfo{year}{1995}).

\bibitem[{\citenamefont{Keefer et~al.}(2001)\citenamefont{Keefer, Gramowski,
  and Gross}}]{Keefer_2001}
\bibinfo{author}{\bibfnamefont{E.~W.} \bibnamefont{Keefer}},
  \bibinfo{author}{\bibfnamefont{A.}~\bibnamefont{Gramowski}},
  \bibnamefont{and} \bibinfo{author}{\bibfnamefont{G.~W.} \bibnamefont{Gross}},
  \bibinfo{journal}{J. Neurophysiol.} \textbf{\bibinfo{volume}{86}},
  \bibinfo{pages}{3030} (\bibinfo{year}{2001}).

\bibitem[{\citenamefont{Haldeman and Beggs}(2005)}]{Haldeman_2005}
\bibinfo{author}{\bibfnamefont{C.}~\bibnamefont{Haldeman}} \bibnamefont{and}
  \bibinfo{author}{\bibfnamefont{J.~M.} \bibnamefont{Beggs}},
  \bibinfo{journal}{Phys. Rev. Lett.} \textbf{\bibinfo{volume}{94}},
  \bibinfo{pages}{058101} (\bibinfo{year}{2005}).

\bibitem[{\citenamefont{Gross and Schwalm}(1994)}]{Gross_1994}
\bibinfo{author}{\bibfnamefont{G.~W.} \bibnamefont{Gross}} \bibnamefont{and}
  \bibinfo{author}{\bibfnamefont{F.~U.} \bibnamefont{Schwalm}},
  \bibinfo{journal}{J. Neurosci. Meth.} \textbf{\bibinfo{volume}{52}},
  \bibinfo{pages}{73} (\bibinfo{year}{1994}).

\bibitem[{\citenamefont{Segev et~al.}(2002)\citenamefont{Segev, Benveniste,
  Hulata, Cohen, Palevski, Kapon, Shapira, and Ben-Jacob}}]{Segev_2002}
\bibinfo{author}{\bibfnamefont{R.}~\bibnamefont{Segev}},
  \bibinfo{author}{\bibfnamefont{M.}~\bibnamefont{Benveniste}},
  \bibinfo{author}{\bibfnamefont{E.}~\bibnamefont{Hulata}},
  \bibinfo{author}{\bibfnamefont{N.}~\bibnamefont{Cohen}},
  \bibinfo{author}{\bibfnamefont{A.}~\bibnamefont{Palevski}},
  \bibinfo{author}{\bibfnamefont{E.}~\bibnamefont{Kapon}},
  \bibinfo{author}{\bibfnamefont{Y.}~\bibnamefont{Shapira}}, \bibnamefont{and}
  \bibinfo{author}{\bibfnamefont{E.}~\bibnamefont{Ben-Jacob}},
  \bibinfo{journal}{Phys. Rev. Lett.} \textbf{\bibinfo{volume}{88}},
  \bibinfo{pages}{118102} (\bibinfo{year}{2002}).

\bibitem[{\citenamefont{Beggs and Plenz}(2003)}]{Beggs_2003}
\bibinfo{author}{\bibfnamefont{J.~M.} \bibnamefont{Beggs}} \bibnamefont{and}
  \bibinfo{author}{\bibfnamefont{D.}~\bibnamefont{Plenz}}, \bibinfo{journal}{J.
  Neurosci.} \textbf{\bibinfo{volume}{23}}, \bibinfo{pages}{11167}
  (\bibinfo{year}{2003}).

\bibitem[{\citenamefont{Beggs and Plenz}(2004)}]{Beggs_2004}
\bibinfo{author}{\bibfnamefont{J.~M.} \bibnamefont{Beggs}} \bibnamefont{and}
  \bibinfo{author}{\bibfnamefont{D.}~\bibnamefont{Plenz}}, \bibinfo{journal}{J.
  Neurosci.} \textbf{\bibinfo{volume}{24}}, \bibinfo{pages}{5216}
  (\bibinfo{year}{2004}).

\bibitem[{\citenamefont{Wagenaar et~al.}(2006)\citenamefont{Wagenaar, Nadasdy,
  and Potter}}]{Wagenaar_2006}
\bibinfo{author}{\bibfnamefont{D.~A.} \bibnamefont{Wagenaar}},
  \bibinfo{author}{\bibfnamefont{Z.}~\bibnamefont{Nadasdy}}, \bibnamefont{and}
  \bibinfo{author}{\bibfnamefont{S.~M.} \bibnamefont{Potter}},
  \bibinfo{journal}{Phys. Rev. E} \textbf{\bibinfo{volume}{73}},
  \bibinfo{pages}{051907} (\bibinfo{year}{2006}).

\bibitem[{\citenamefont{Ham et~al.}(2007)\citenamefont{Ham, Bettencourt, Gross,
  and McDaniel}}]{Ham_2007}
\bibinfo{author}{\bibfnamefont{M.~I.} \bibnamefont{Ham}},
  \bibinfo{author}{\bibfnamefont{L.~M.~A.} \bibnamefont{Bettencourt}},
  \bibinfo{author}{\bibfnamefont{G.~W.} \bibnamefont{Gross}}, \bibnamefont{and}
  \bibinfo{author}{\bibfnamefont{F.~D.} \bibnamefont{McDaniel}},
  \bibinfo{journal}{to appear in J. Comp. Neurosci.}  (\bibinfo{year}{2007}).

\bibitem[{\citenamefont{Jia et~al.}(2004)\citenamefont{Jia, Sano, Lai, and
  Chan}}]{Jia_2004}
\bibinfo{author}{\bibfnamefont{L.~C.} \bibnamefont{Jia}},
  \bibinfo{author}{\bibfnamefont{M.}~\bibnamefont{Sano}},
  \bibinfo{author}{\bibfnamefont{P.-Y.} \bibnamefont{Lai}}, \bibnamefont{and}
  \bibinfo{author}{\bibfnamefont{C.~K.} \bibnamefont{Chan}},
  \bibinfo{journal}{Phys. Rev. Lett.} \textbf{\bibinfo{volume}{93}},
  \bibinfo{pages}{088101} (\bibinfo{year}{2004}).

\bibitem[{\citenamefont{Jimbo et~al.}(1999)\citenamefont{Jimbo, Tateno, and
  Robinson}}]{Jimbo_1999}
\bibinfo{author}{\bibfnamefont{Y.}~\bibnamefont{Jimbo}},
  \bibinfo{author}{\bibfnamefont{T.}~\bibnamefont{Tateno}}, \bibnamefont{and}
  \bibinfo{author}{\bibfnamefont{H.~P.~C.} \bibnamefont{Robinson}},
  \bibinfo{journal}{Biophys. J.} \textbf{\bibinfo{volume}{76}},
  \bibinfo{pages}{670} (\bibinfo{year}{1999}).

\bibitem[{\citenamefont{Marom and Shahaf}(2002)}]{Marom_2002}
\bibinfo{author}{\bibfnamefont{S.}~\bibnamefont{Marom}} \bibnamefont{and}
  \bibinfo{author}{\bibfnamefont{G.}~\bibnamefont{Shahaf}},
  \bibinfo{journal}{Q. Rev. Biophys.} \textbf{\bibinfo{volume}{35}},
  \bibinfo{pages}{63} (\bibinfo{year}{2002}).

\bibitem[{\citenamefont{DeMarse et~al.}(2001)\citenamefont{DeMarse, Wagenaar,
  Blau, and Potter}}]{Demarse_2001}
\bibinfo{author}{\bibfnamefont{T.~B.} \bibnamefont{DeMarse}},
  \bibinfo{author}{\bibfnamefont{D.~A.} \bibnamefont{Wagenaar}},
  \bibinfo{author}{\bibfnamefont{A.~W.} \bibnamefont{Blau}}, \bibnamefont{and}
  \bibinfo{author}{\bibfnamefont{S.~M.} \bibnamefont{Potter}},
  \bibinfo{journal}{Auton. Rob.} \textbf{\bibinfo{volume}{11}},
  \bibinfo{pages}{305} (\bibinfo{year}{2001}).

\bibitem[{\citenamefont{Tateno et~al.}(2002)\citenamefont{Tateno, Kawana, and
  Jimbo}}]{Tateno_2002}
\bibinfo{author}{\bibfnamefont{T.}~\bibnamefont{Tateno}},
  \bibinfo{author}{\bibfnamefont{A.}~\bibnamefont{Kawana}}, \bibnamefont{and}
  \bibinfo{author}{\bibfnamefont{Y.}~\bibnamefont{Jimbo}},
  \bibinfo{journal}{Phys. Rev. E} \textbf{\bibinfo{volume}{65}},
  \bibinfo{pages}{051924} (\bibinfo{year}{2002}).

\bibitem[{\citenamefont{Segev et~al.}(2001)\citenamefont{Segev, Shapira,
  Benveniste, and Ben-Jacob}}]{Segev_2001}
\bibinfo{author}{\bibfnamefont{R.}~\bibnamefont{Segev}},
  \bibinfo{author}{\bibfnamefont{Y.}~\bibnamefont{Shapira}},
  \bibinfo{author}{\bibfnamefont{M.}~\bibnamefont{Benveniste}},
  \bibnamefont{and}
  \bibinfo{author}{\bibfnamefont{E.}~\bibnamefont{Ben-Jacob}},
  \bibinfo{journal}{Phys. Rev. E} \textbf{\bibinfo{volume}{64}},
  \bibinfo{pages}{011920} (\bibinfo{year}{2001}).

\bibitem[{\citenamefont{Schneidman et~al.}(2006)\citenamefont{Schneidman,
  {Berry II}, Segev, and Bialek}}]{Schneidman_2006}
\bibinfo{author}{\bibfnamefont{E.}~\bibnamefont{Schneidman}},
  \bibinfo{author}{\bibfnamefont{M.~J.} \bibnamefont{{Berry II}}},
  \bibinfo{author}{\bibfnamefont{R.}~\bibnamefont{Segev}}, \bibnamefont{and}
  \bibinfo{author}{\bibfnamefont{W.}~\bibnamefont{Bialek}},
  \bibinfo{journal}{Nature} \textbf{\bibinfo{volume}{440}},
  \bibinfo{pages}{1007} (\bibinfo{year}{2006}).

\end{thebibliography}
\end{document}